\documentclass[aps,12pt]{revtex4}

\usepackage{epsfig}
\usepackage{graphicx}
\usepackage{amsmath,amssymb}
\usepackage{color}

\parskip=\medskipamount

\newcommand{\eq}[1]{(\ref{#1})}
\newcommand{\fig}[1]{Fig. \ref{#1}}

\newcommand{\be}{\begin{equation}}
\newcommand{\ee}{\end{equation}}

\newcommand\disp{\displaystyle}

\newcommand\eps{\varepsilon}

\newcommand{\la}{\left<}
\newcommand{\ra}{\right>}
\newcommand{\appropto}{\mathrel{\vcenter{
  \offinterlineskip\halign{\hfil$##$\cr
    \propto\cr\noalign{\kern2pt}\sim\cr\noalign{\kern-2pt}}}}}

\begin{document}

\title{Anomalous 1D fluctuations of a simple 2D random walk in a large deviation regime}

\author{Sergei Nechaev$^{1,2}$, Kirill Polovnikov$^{3,4}$, Senya Shlosman$^{4,5,6}$, Alexander Valov$^{7}$, and Alexander Vladimirov$^{5}$}

\affiliation{$^1$ Interdisciplinary Scientific Center Poncelet, CNRS UMI 2615, 119002 Moscow, Russia \\ $^2$ P.N. Lebedev Physical Institute RAS, 119991 Moscow, Russia \\ $^3$ Physics Department, Lomonosov Moscow State University, 119992 Moscow, Russia \\ $^4$ Skolkovo Institute of Science and Technology, 143005 Skolkovo, Russia \\ $^5$ Institute of Information Transmission Problems RAS, 127051 Moscow, Russia \\ $^6$ Aix-Marseille University, Universite of Toulon, CNRS, CPT UMR 7332, 13288, Marseille, France \\ $^7$ N.N. Semenov Institute of Chemical Physics RAS, 119991 Moscow, Russia}

\begin{abstract}

The following question is the subject of our work: could a two-dimensional random path pushed by some constraints to an improbable "large deviation regime", possess extreme statistics with one-dimensional Kardar-Parisi-Zhang (KPZ) fluctuations? The answer is positive, though non-universal, since the fluctuations depend on the underlying geometry. We consider in details two examples of 2D systems for which imposed external constraints force the underlying stationary stochastic process to stay in an atypical regime with anomalous statistics. The first example deals with the fluctuations of a stretched 2D random walk above a semicircle or a triangle. In the second example we consider a 2D biased random walk along a channel with forbidden voids of circular and triangular shapes. In both cases we are interested in the dependence of a typical span $\la d(t) \ra \sim t^{\gamma}$ of the trajectory of $t$ steps above the top of the semicircle or the triangle. We show that $\gamma = \frac{1}{3}$, i.e. $\la d(t) \ra$ shares the KPZ statistics for the semicircle, while $\gamma=0$ for the triangle. We propose heuristic derivations of scaling exponents $\gamma$ for different geometries, justify them by explicit analytic computations and compare with numeric simulations. For practical purposes, our results demonstrate that the geometry of voids in a channel might have a crucial impact on the width of the boundary layer and, thus, on the heat transfer in the channel.

\end{abstract}

\maketitle

\section{Introduction}

Intensive investigation of extremal problems of correlated random variables in statistical mechanics has eventually led mathematicians, and then, physicists, to understanding that the Gaussian distribution is not as ubiquitous in nature, as it has been thought over the centuries, and shares its omnipresence (at least in one dimension) with another distribution, known as the Tracy-Widom (TW) law. The necessary (though not sufficient) feature of the TW distribution is the width of the distribution, controlled by the critical exponent $\nu=\frac{1}{3}$, the so-called Kardar-Parisi-Zhang (KPZ) exponent. For the first time, the KPZ exponent has appeared in the seminal paper \cite{kpz} (see \cite{halpin} for review) as the growth exponent in a non-equilibrium one-dimensional directed stochastic process, for which the theoretical analysis has been focused mainly on statistical properties of the enveloping surface developing in time.

Nowadays one has accumulated many examples of one-dimensional statistical systems of seemingly different physical nature, whose fluctuations are controlled by the KPZ exponent $\gamma=\frac{1}{3}$, contrary to the exponent $\gamma=\frac{1}{2}$ typical for the distribution of independent random variables. Among such examples it is worth mentioning the restricted solid-on-solid \cite{rsos} and Eden \cite{eden} models, molecular beam epitaxy \cite{MBE}, polynuclear growth \cite{M,PS,PNG,BR, J}, several ramifications of the ballistic deposition \cite{Mand,MRSB,KM,BMW}, alignment of random sequences \cite{align}, traffic models of TASEP type \cite{tasep}, (1+1)D vicious walks \cite{vw}, area-tilted random walks \cite{ISV}, and 1D directed polymer in random environment \cite{dotsenko}. Recently, this list has been replenished by the one-dimensional modes describing the fluctuational statistics of cold atoms \cite{cold}.

Here we study a two-dimensional model demonstrating the one-dimensional KPZ critical behavior.
The interest to such systems is inspired by the (1+1)D model proposed by H. Spohn and P. Ferrari in \cite{ferrari} where they discussed the statistics of 1D directed random walks evading the semicircle. As the authors stated in \cite{ferrari}, their motivation was as follows. It is known that the fluctuations of a top line in a bunch of $n$ one-dimensional directed "vicious walks" glued at their extremities (ensemble of world lines of free fermions in 1D) are governed by the Tracy-Widom distribution \cite{vw}. Proceeding as in \cite{spohn-pr-fer}, define the averaged position of the top line and look at its fluctuations. In such a description, all vicious walks lying below the top line, play a role of a "mean field" of the "bulk", pushing the top line to some equilibrium position. Fluctuations around this position are different from fluctuations of a free random walk in absence of the "bulk". Replacing the effect of the "bulk" by the semicircle, one arrives at the Spohn-Ferrari model where the 1D directed random walk stays above the semicircle, and its interior is inaccessible for the path. In \cite{ferrari} the authors confirmed that this system has a KPZ critical exponent.

In our work we study fluctuations of a two-dimensional random path pushed by some geometric constraints to an improbable "large deviation regime" and ask the question whether it could possess extreme statistics with one-dimensional Kardar-Parisi-Zhang (KPZ) fluctuations. We propose the "minimal" model and in its frameworks formulate the answer to the question posed above.

We consider an ensemble of two-dimensional random paths stretched over some forbidden void with prescribed geometry and characteristic scale, $R$. Stretching is induced by the restriction on wandering times, $t$, such that $c R < t \ll R^2$. The resulting paths conformations are "atypical" since their realizations would be highly improbable in the ensemble of unconstrained trajectories which exhibit the Gaussian behavior. Statistics in such a tiny subset of the Gaussian ensemble is naturally controlled by collective behavior of strongly correlated modes, thus, for some geometries one might expect extreme distribution with KPZ scaling for fluctuations, similarly to the (1+1)D model of \cite{ferrari}. Simple dimensional analysis supports this hypothesis. Indeed, consider a realization of the stretched random walk in 2D with the diffusion coefficient $D$ evading a circular void in two distinct regimes. An unconstrained $t$-step random walk, with $t \gg R^2$ fluctuates freely and does not feel the constraint, thus, the only possible combination of $D$ and $t$, which has the dimension of length, could be $d \sim (Dt)^{1/2}$ for the typical span of the path. In the opposite regime, $\pi R< t \ll R^2$, the chain statistics is essentially perturbed by the constraint. In the limit of strong stretching, $t\sim R$, these two parameters ($t$ and $R$) should enter symmetrically in the combination for the span. The suitable dimension is given by the scaling expression $d \sim (D R t)^{\gamma}$ with $\gamma = 1/3$, which is the unique combination that in the limit $t \ll R^2$ recovers a physically relevant condition $d \ll R$  and at $t\sim R^2$ gives $d\sim R$. Such a dimensional analysis strongly relies on the uniqueness of the scale, characterizing constraint, which is true only for homogeneously curved boundaries and breaks down for more complex algebraic curves, like cubic parabola or boundaries with a local cusp (triangle). In particular, trajectories above triangular obstacles fluctuate irrespectively to the size of the void even in the "strong stretching regime".

The paper is organized as follows. in Section II we formulate the model of a 2D stretched random walk above the semicircle (model "S") and the triangle (model "T") and provide scaling arguments for the averaged span of paths above the top of these voids, supported by numeric simulations. in Section III we solve the diffusion equation in 2D in the limit of stretched trajectories $N = cR$ above the semicircle and the triangle. in Section IV we discuss the results of numeric simulations for fluctuations of biased 2D random walks above forbidden voids of different shapes. in Section V we summarise the obtained results and discuss their possible generalizations and applications.

\section{Two-dimensional random walk stretched over the voids of various shapes}

\subsection{The model}

\begin{figure}[ht]
\centerline{\includegraphics[width=16cm]{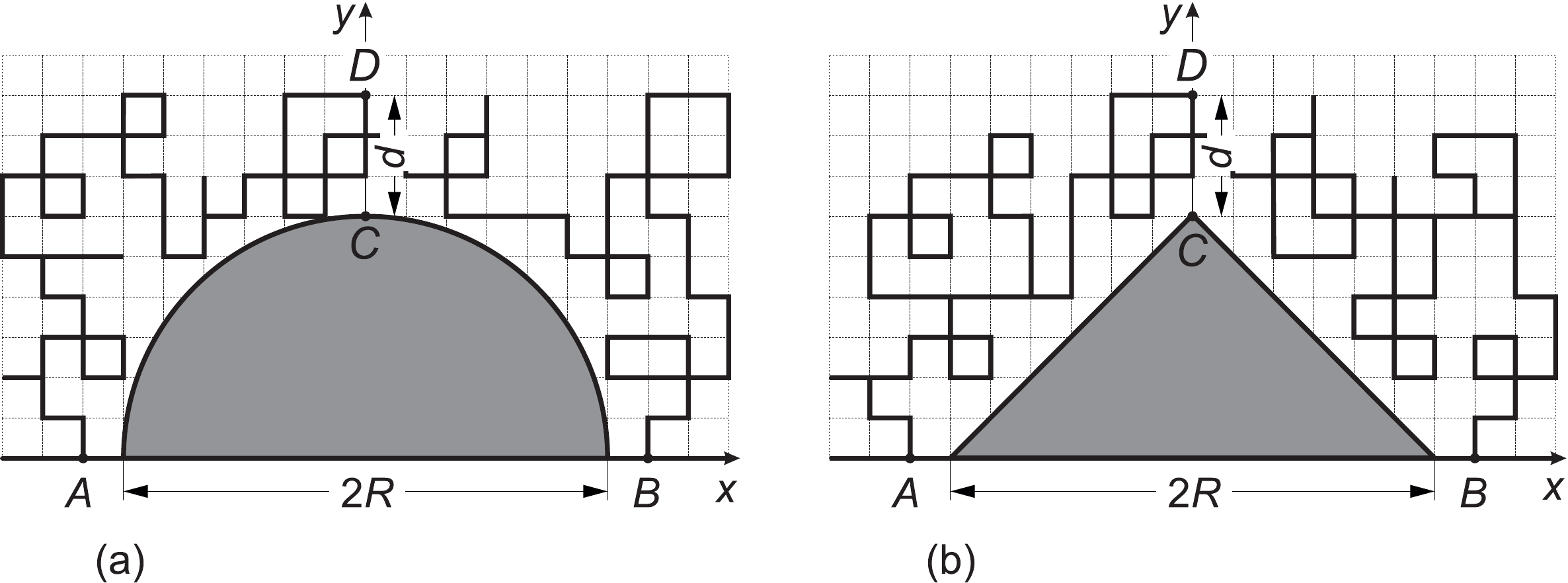}}
\caption{Two-dimensional random walk on a square lattice in the upper half-plane, that evades: (a) Model "S": the semicircle of radius $R$, and (b)  Model "T": the rectangular triangle of base $2R$. The number of steps $N \ll R^2$.}
\label{fig:01}
\end{figure}

We begin with the lattice version of the model. Consider the $N$-step symmetric random walk, $\mathbf{r}_n=\{x_n,y_n\}$, on a two-dimensional square lattice in a discrete time $n$ ($n=1,2,...,N$). The walk begins at the point $A$, terminates after $N$ steps at the point $B$, and satisfies three requirements: (i) for any $n$ one has $y_n\ge 0$, (ii) the random walk evades the semicircle of the diameter $2R$, or the rectangular triangle of the base $2R$, i.e. it remains outside  the obstacles shown in \fig{fig:01} and (iii) the total number of steps is much less than the squared size of the obstacle, $N \ll R^2$. Note that the requirement (i) is not crucial and can be easily relaxed. The points $A$ and $B$ are located in one lattice spacing from left and right extremities of the obstacle (semicircle or triangle) -- see \fig{fig:01}.

We are interested in the critical exponents $\gamma$ of in the dependence $\la d(R)\ra \sim R^{\gamma}$ as $R\to \infty$ for the model "S" and the model "T". In this section we provide qualitative scaling estimates for the mean span of two-dimensional stretched paths above any smooth algebraic curve and support our analysis by numeric simulations.

\subsection{Scaling arguments: from semicircle to algebraic curve}

Normally, a stretched path follows the straight line as much as possible, and gets curved only if curving cannot be avoided. A random path which has to travel a horizontal distance, $x_S$, is localized within a strip of typical width ("span" in a vertical direction), $y_S \sim\sqrt{x_S}$. If the path is forced to travel a distance $x_S$ along some curved arc, and the arc fits this strip, the curving of the arc can be ignored. Consider a path that has to follow a circle of radius $R$. Note that the arc of that circle of length $x_S$ fits a strip of width $x_S^2/R$. Therefore the arc length, curving of which can be ignored, is
\be
x_S^2/R\leq\sqrt{x_S}
\ee
This puts a limit to $x_S$: it has to be at most $R^{2/3}$. At shorter distances the stretched path can be considered as an unconstrained random walk. Therefore, the span in vertical direction is of the order of $y_S \sim \sqrt{R^{2/3}}=R^{1/3}$. Beyond this "blob" of length $x_S=R^{2/3}$ the arc itself deviates considerably from a straight segment, and the estimate $\sqrt{x_S}$ for fluctuations above it is no longer applicable.

\begin{figure}[ht]
\centerline{\includegraphics[width=14cm]{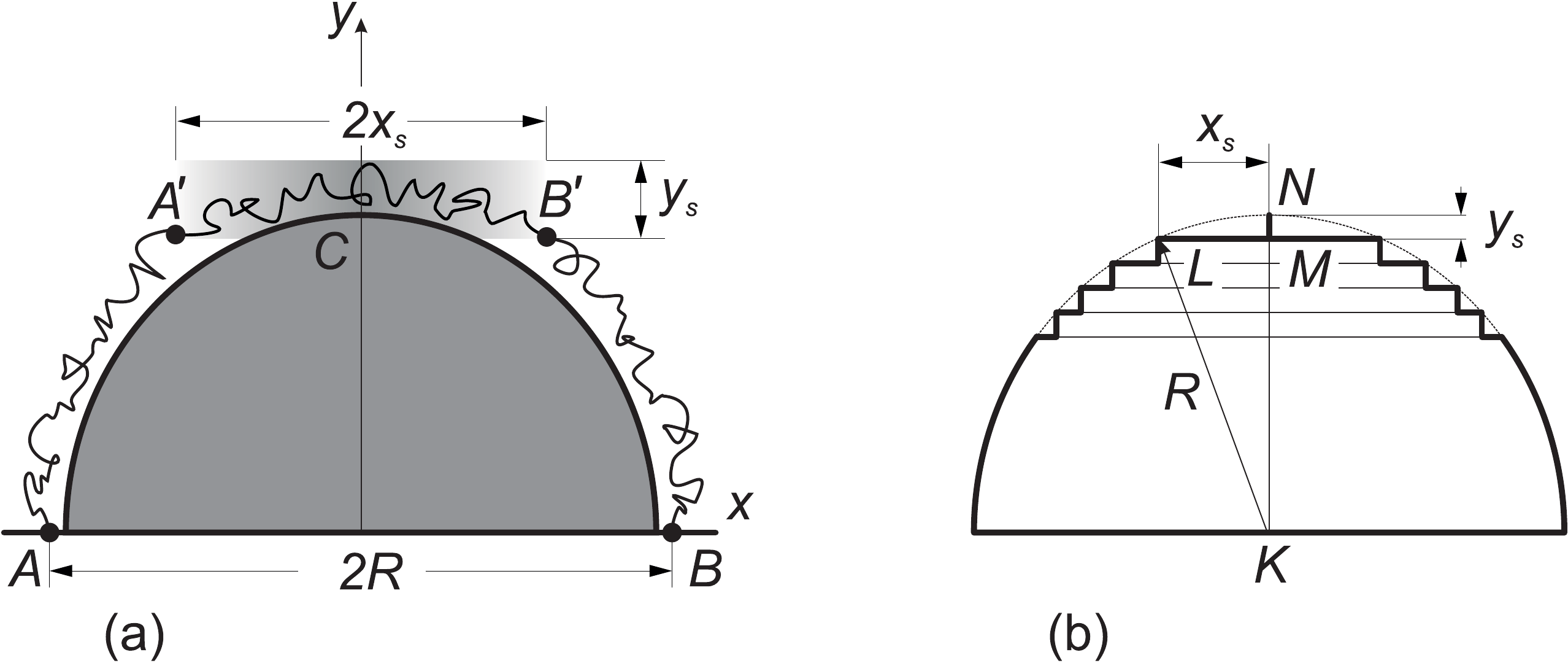}}
\caption{(a) Two-dimensional random walk evading the semicircle. The part $A'B'$ lies above the essentially flat region of the semicircle. The figure (b) provides an auxiliary geometric construction for Eq.\eq{st}.}
\label{fig:02}
\end{figure}

To add some geometric flavor to these arguments, consider \fig{fig:02}a and denote by $y_S$ an average span of the path in vertical direction above the point $C$ of the semicircle, and by $x_S$ -- the typical size of the horizontal segment, along which the semicircle can be considered as nearly flat. We divide the path in three parts: $AA'$, $A'B'$ and $B'B$. The parts $AA'$ and $BB'$ of the trajectory run above essentially curved domains, while the part $A'B'$ constitutes a segment that is mainly flat. Schematically this is shown in \fig{fig:02}b: in the limit $y_S \ll R$, the horizontal segment $LM$ linearly approximates the corresponding arc of the circle. Our goal is to estimate $x_S$ and to provide self-consistent scaling arguments for fluctuations $y_S(R)\sim R^{\gamma}$ of the stretched path.

From the triangle $KLM$ we have:
\be
|LM| = \sqrt{R^2 - |KM|^2} = \sqrt{R^2 - (R-y_S)^2}\,\Big|_{y_S\ll R} \approx \sqrt{2Ry_S}
\label{st}
\ee
Since $|LM|\equiv x_S$, the condition of stretched trajectories, $y_S\ll R$, implies the relation
\be
x_S\sim \sqrt {R y_S}
\label{st1}
\ee
Consider now a two-dimensional random walk which starts at the point $L$ near the left extremity of the excluded shape and terminates anywhere at the segment $MN$ ($|MN|\equiv y_S$). Since the horizontal support, $|LM|=x_S$, of the path is flat, the span of the trajectory in vertical direction is the same as for an ordinary random walk. Thus, we can estimate the typical span, $y_S$, as
\be
y_S\sim \sqrt{x_S}
\label{yx}
\ee
On the scales larger than $x_S$ the curvature of the semicircle becomes essential and the relation \eq{st1} is not valid anymore.

It should be noted that \eq{yx} is insensitive to a specific way of stretching. Eq \eq{yx} remains unchanged even if we introduce an asymmetry in random jumps along $x$--axis while keeping the symmetry of jumps in $y$ direction. Substituting the scaling \eq{yx} into \eq{st}, we obtain for the semicircle (the model "S"):
\be
x_S \sim \sqrt{R \sqrt{x_S}}
\label{st2}
\ee
From the first equation of \eq{st2} we get for the semicircle:
\be
x_S \sim R^{2/3}; \quad y_S \sim \sqrt{x_S} \sim R^{1/3}
\label{semicircle}
\ee
which implies that $\gamma=\frac{1}{3}$. The analytic computations presented in Section III for the model "S" support this conclusion. Let us note that the large-scale deviation principle for the  constrained 1D random walk process has been discussed recently in \cite{baruch}.

We expect that our scaling can be extended to random walks above any algebraic curve. The critical exponent $\gamma$ for the fluctuations of the stretched random walk above the curve $\Gamma$: $y=x^{\eta}$ in 2D should be understood as follows. Define the characteristic length scale, $R$, and represent the curve $\Gamma$ in dimensionless units:
\be
\frac{y}{R} \approx \left(\frac{x}{R}\right)^{\eta}
\label{gamma}
\ee
For $\eta=2$ we are back to semicircle \eq{st1}. As in the former case, Eq. \eq{gamma} should be equipped by \eq{yx}. Solving these equations self-consistently, we get the following scaling dependence for the span $y_G(R)$ of the path above the curve $\Gamma$:
\be
y_G(R) \sim R^{\gamma}; \qquad  \gamma = \frac{\eta-1}{2\eta-1}
\label{gamma2}
\ee
Note that for $\eta \to\infty$ the curve is straight and we get the fluctuations with the standard Gaussian exponent, $\gamma = 1/2$, which is the exponent of fluctuations above the straight line. The opposite case of a cusp can be approached in the limit $\eta \to 1$, which gives $\gamma = 0$. This result rhymes well with simulations of paths stretched over the triangle (see below) and analytic solution of the diffusion equation (Section III).

\subsection{Heuristic arguments: triangle}

\begin{figure}[ht]
\centerline{\includegraphics[width=16cm]{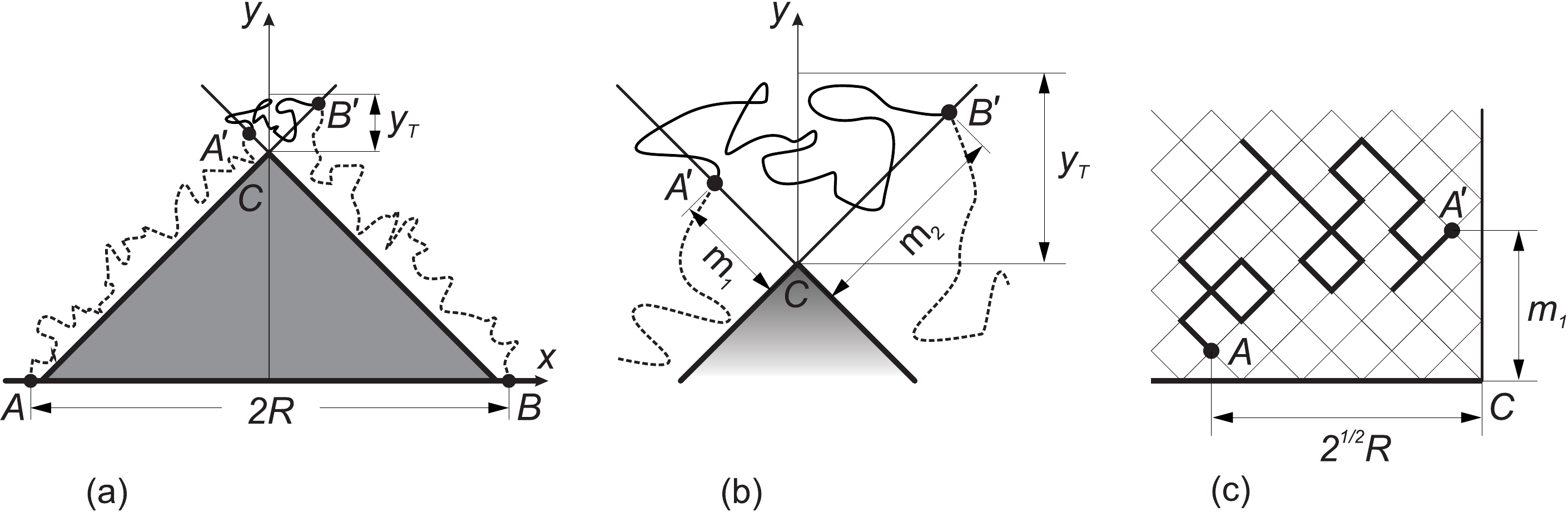}}
\caption{(a) Two-dimensional random walk evading a triangle, and (b) the magnified part of the system near the tip of the triangle. The points $A'$ and $B'$ are respectively the points of the first entry by the random walk into the wedge above the point $O$ and the last exit from it; (c) subpart of the random walk from $A$ to $A'$ which does not escape the wedge with zero's boundary conditions.}
\label{fig:03}
\end{figure}

To estimate the fluctuations of the path of $N$ steps stretched over the triangle of base $2R$, the above arguments for the semicircle need to be modified since the curvature of the triangle is non-analytic being concentrated at one single point $C$ at the tip of the obstacle. To proceed, some auxiliary construction should be used -- see \fig{fig:03}a and its zoom in \fig{fig:03}b.

We split the full trajectory between points $A$ and $B$ into three parts: the part of $N_1$ steps running between point $A$ and first entry to the point $A'$, the part of $M$ steps running between the points $A'$ and $B'$, and the part $N_2$ running between $B$ and first entry to the point $B'$. The parts $N_1$ and $N_2$ lie above the flat boundaries of the triangle $AOB$, while the part $A'B'$ is located in the vicinity of the tip of the triangle. The partition function, $Z_N$, of the full $N$-step path with the extremities at $A$ and $B$ can be written as follows:
\be
Z_N(R) = \sum_{\{N_1+M+N_2=N\}} \hspace{1mm}  \sum_{\{m_1,m_2\}} U_{N_1}(m_1, R)\, W_{M}(m_1, m_2)\, U_{N_2}(m_2, R)
\label{triang1}
\ee
where $U_{N_1}(m_1,R)$, $W_{M}(m_1,m_2)$, $U_{N_2}(m_2,R)$ are, respectively, the partition functions of parts $AA'$, $A'B'$ and $B'B$, the first sum runs over $N_1,M,N_2$ such that $N_1+M+N_2=N$ and $m_1$ and $m_2$ are the positions of the points $A'$ and $B'$ at the edges of the wedge (see \fig{fig:03}b). The partition functions $U_{N_i}(m_i,R)$  ($i=1,2$) can be computed on the lattice in the geometry shown in \fig{fig:03}c with zero's boundary conditions in the wedge
\be
U_{N_i}(m_i,R) = \frac{1}{\pi^2}\int_0^{\pi} d q_1 \int_0^{\pi} dq_2 \sin(q_1 R\sqrt{2})\sin q_1 \sin(q_2 m_i) \sin q_2 (\cos q_1 + \cos q_2)^{N_i}
\label{eq:ray}
\ee
where $q_1$ and $q_2$ are the Fourier-transformed coordinates along the wedge sides. In \eq{eq:ray} the subpath of $N_i$ steps is not yet stretched, i.e. $N_i$, $m_i$ and $R$ are all independent.

Our goal now is to estimate the typical length $M$ of the subpath between the points $A'$ and $B'$ as shown in \fig{fig:03}b. Below we show that $M = \mathrm{const}$ which immediately leads to the conclusion that $y_T=\mathrm{const}$. To proceed, it is convenient to pass to the grand canonical formulation of the problem. Let us define the generating function $Z(s,R) = \sum_{N=0}^{\infty} Z_N(R) s^N$ of the grand canonical ensemble, and introduce the variable $\beta = -\ln s$, which has the sense of an "energy" attributed to each step of the trajectory (note that $\beta>0$ since $0<s<1$). To "stretch" the trajectory, we should imply $\beta\gg 1$. In the stretched regime $\beta\gg 1$ the generating function of $U_{N_i}(m_i,R)$ can be estimated as follows
\be
U(\beta,m_i,R) = \int_{0}^{\infty} U_{N_i}(m_i,R) e^{-\beta N_i} d N_i \sim \frac{m_i R\, \beta_s^{3/4}\exp\left(-2\sqrt{\beta_s}\,\sqrt{m_i^2+2R^2} \right)} {(m_i^2+2R^2)^{5/4}}
\label{eq:U}
\ee
where we also supposed that $R\gg 1$ and introduced $\beta_s = \beta - \ln 4$. The shift by $\ln 4$ in $\beta$ comes from the fact that the partition function \eq{eq:ray} on the square lattice has the exponential prefactor $4^{N_i} \equiv e^{N_i\ln 4}$ which should be properly taken into account in the generation function.

The generation function of $Z_N(R)$ reads:
\be
Z(\beta,R) = \sum_{N=0}^{\infty} Z_N(R) s^N = \sum_{\{m_1,m_2\}} U(\beta, m_1,R)\, W(\beta, m_1,m_2)\, U(\beta, m_2,R)
\label{triang2}
\ee
Now we should account for the contribution of $W(\beta,m_1,m_2)$ to \eq{triang2}. Note, that each step of the path of length $M$ between points $A'$ and $B'$ carries the energy $\beta>0$. To maximize the contribution of $W(\beta,m_1,m_2)$, one should make the corresponding length $M$ between $A'$ and $B'$ as small as possible, since we loose the energy $\beta M$ for $M$ steps. Thus, $M$ should be of order of $\max(m_1,m_2)$. From \eq{eq:U}--\eq{triang2} we immediately conclude that at $\beta_s\gg 1$ the major contribution to $Z(\beta)$ comes from $m_i$ which should be as small as possible, i.e. $m_1\sim m_2 = \mathrm{const}$. This immediately implies that $M=\mathrm{const}$ and the span $y_T$ (for $N=cR$ and $R\gg 1$) becomes independent on $R$:
\be
y_T = \mathrm{const}
\label{eq:triang4}
\ee
The same conclusion follows from the solution of the boundary problem in the open wedge for the model "T" -- see Section IV. Note, that putting $\eta = 1$ into \eq{gamma2}, we get $\gamma=0$, thus arriving at the same conclusion of independence of the span of fluctuations of stretched path above the tip of the triangle on $R$.

\subsection{Numerics}

Here we confirm our scaling and heuristic analyses of the mean height of the 2D ensemble of stretched trajectories above the top of the semicircle and the triangle using numeric simulations. Let us emphasize that this part pursues mainly the illustrative goals, while detailed analytic computations for distribution functions are provided in the following Section III.

Specifically, we have enumerated all $N$-step paths on the square lattice, travelling from the point $A(-R-1,0)$ to the point $D(0,R+d)$ above the top of the semicircle or triangle, as shown in \fig{fig:01}a,b. Let us emphasize that this is an exact path counting problem. The step length of a path coincides with the lattice spacing. We allow all steps: "up", "down","right", "left" and set the constraint $N = cR$ on the total number of steps. The values of $R$ and $c$ in the simulations are as follows: $R=\{10, 20, 40, 60, 100, 200, 300, 400\}$ and $c=\{5, 10, 20\}$. Counting ensemble of trajectories from $A$ to $D$ is sufficient for extracting the scaling dependence $\la d(R)\ra \sim R^{\gamma}$ since the part of the path from $A$ to $C$ is independent from the part from $C$ to $B$. The enumeration of trajectories respects boundary conditions and is performed recursively within the box of size $3R \times 3R$ with the bottom left corner located at the point $(-2R, 0)$.

\begin{figure}[ht]
\centerline{\includegraphics[width=16cm]{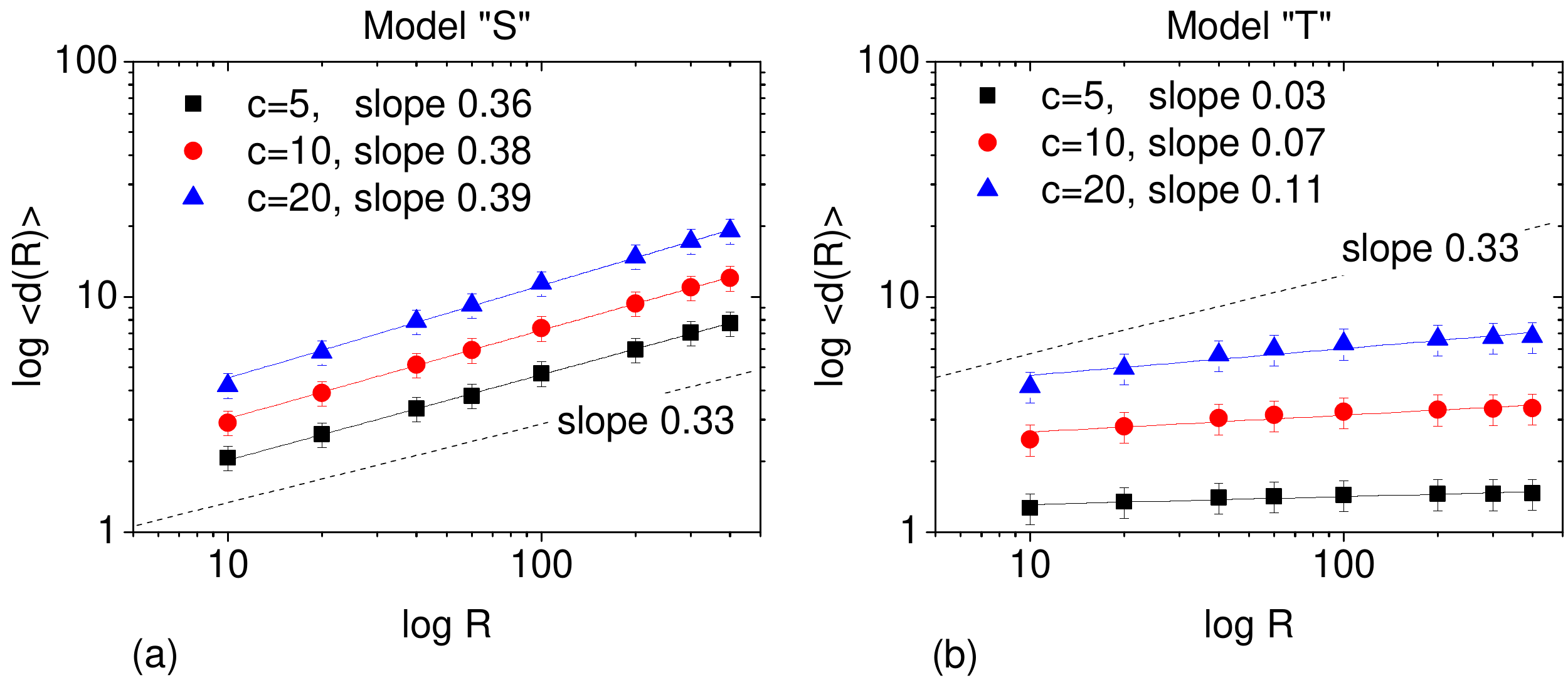}}
\caption{The mean deviation of the path of $N$ steps above the semicircle (a) and the triangle (b) for different values of the parameter $c$, which controls "stretching" of the path (the less $c$ the more stretched the path).}
\label{fig:04}
\end{figure}

The results of simulations in doubly-logarithmic scale $\log \la d(R) \ra $ \emph{vs} $\log R$ for the averaged span $\la d \ra$ of paths above the top of the semicircle of radius $R$ and the triangle of base $2R$ are presented in \fig{fig:04}. The physical meaning of the constant $c$ is the effective "stretching" of the path: the less $c$, the more stretched the path (definitely, on the square lattice $c>4$.

As one sees from \fig{fig:04}a, all stretched paths above the semicircle demonstrate the scaling $\la d(R) \ra \sim R^{\gamma}$ with the exponent $\gamma$ close to $1/3$. For less stretched paths (larger values of $c$) the deviation form the scaling with $\gamma=\frac{1}{3}$ becomes notable. The span of stretched 2D trajectories above the tip of the triangle shown in \fig{fig:04}b are almost independent on $R$ (i.e. the exponent $\gamma$ is close to 0). This result is consistent with our scaling estimates, as well as with the theoretical arguments presented below. Some conjectures about possible physical consequences of the difference between fluctuations of stretched random trajectories above the semicircle and above the triangle are formulated in Section IV.

\section{2D stretched random walks above the semicircle and triangle: analytic results}

\subsection{Semicircle}

The symmetric two-dimensional random walk on a lattice depicted in \fig{fig:01}a in the limit $N\to \infty$, $a\to 0$ (where $a$ is the lattice spacing) where $Na=t$, converges to the two-dimensional Brownian motion of time $t$ with diffusion coefficient $D=\frac{a^2}{4}$, that evades the semicircular void of radius $R$. Let $P(\rho,\phi; \rho_0,\phi_0;t)$ be the probability density to find the random walk of length (time) $t$ at the point $(\rho, \phi)$ above the void under the condition that the path begins at the point $(\rho_0, \phi_0)$. The function $P(\rho,\phi; \rho_0,\phi_0;t) \equiv P(\rho,\phi,t)$ satisfies the diffusion equation in polar coordinates
\be
\begin{cases}
\disp \frac{\partial P(\rho,\phi,t)}{\partial t} = D \left[\frac{1}{\rho}\frac{\partial}{\partial \rho}\left(\rho \frac{\partial P(\rho,\phi,t)}{\partial \rho}\right) + \frac{1}{\rho^2} \frac{\partial^2 P(\rho,\phi,t)}{\partial \phi^2} \right] \medskip \\
P(\rho=R,\phi,t)=P(\rho\to \infty,\phi,t)=P(\rho,\phi=0,t)=P(\rho,\phi=\pi,t)=0 \medskip \\
P(\rho,\phi,0)=\delta(\rho-\rho_0)\delta(\phi-\phi_0)
\end{cases}
\label{eq:s1}
\ee

The explicit solution of \eq{eq:s1} reads
\be
P(\rho,\phi,t)=\sum_{k=1}^{\infty} \frac{2\rho_0}{\pi}\sin(k \phi_0)\sin(k\phi) \int_{0}^{\infty} e^{-\lambda^2 Dt} Z_k(\lambda\rho,\lambda R) Z_k(\lambda\rho_0,\lambda R) \lambda d\lambda
\label{eq:s2}
\ee
where
\be
Z_k(\lambda\rho, \lambda R)=\frac{-J_k(\lambda\rho)N_k(\lambda R)+J_k(\lambda R) N_k(\lambda\rho)}{\sqrt{J^2_k(\lambda R)+N^2_k(\lambda R)}}
\label{eq:s3}
\ee
and $J$ and $N$ denote correspondingly the Bessel and the Neumann functions. Introducing the new variables, $\mu$ and $r$, and making in \eq{eq:s3} the substitution
\be
\lambda=\frac{\mu}{R}, \qquad \rho=R+r,
\label{eq:s3-1}
\ee
we arrive at the following expression for $P(\rho,\phi,t)$:
\be
P(r,\phi,t)=\frac{2\rho_0}{\pi R^2}\sum_{k=1}^{\infty} \sin(k \phi_0)\sin(k\phi)
\int_{0}^{\infty}e^{-\frac{\mu^2 Dt}{R^2}} Z_k\left(\mu+\frac{\mu r}{R},\mu\right) Z_k\left(\mu+\frac{\mu r_0}{R},\mu\right) \mu d\mu
\label{eq:s4}
\ee

The probability to stay above the top of the semicircle consists of two parts: the probability $P'=P\left(r,\phi=\frac{\pi}{2}, t'\right)$ to run from the point $A$ to the point $(r,\phi=\frac{\pi}{2})$ during the time $t'$ and the probability $P''=P\left(r, \phi=\frac{\pi}{2}, t''\right)$ to run from the point $(r,\phi=\frac{\pi}{2})$ to the point $B$ during the time $t''=t-t'$. Obviously, $P'$ and $P''$ are independent, thus the total probability to find path at the point $(r,\phi=\frac{\pi}{2})$ above the semicircle can be estimated as $Q = P'\times P''$ where $t'=t''=t/2$, namely
\be
Q\left(r,\phi=\frac{\pi}{2}, t\right) = \frac{1}{\cal N} P^2\left(r,\phi=\frac{\pi}{2}, t=c R\right); \qquad {\cal N} = \int_{0}^{\infty} P^2\left(r,\phi=\frac{\pi}{2}, t\right) dr
\label{eq:s5}
\ee

\begin{figure}[ht]
\centerline{\includegraphics[width=16cm]{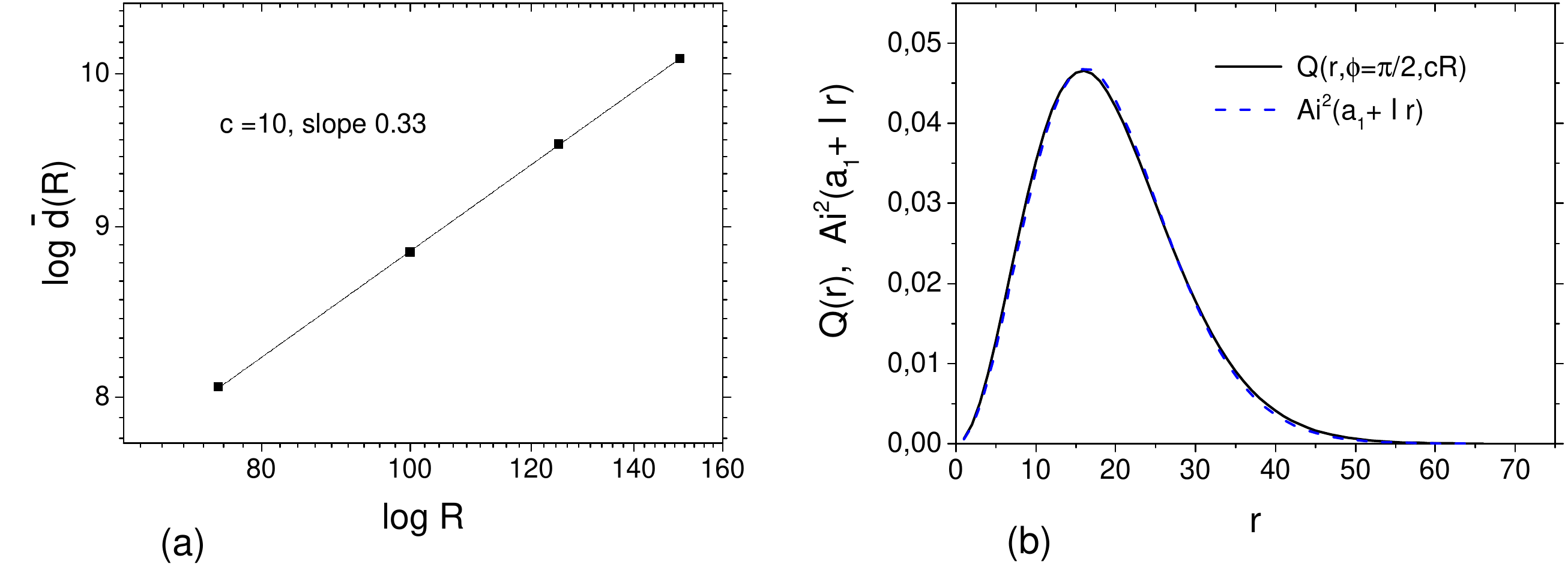}}
\caption{(a) Expectation $\bar{d}$ as a function of $R$ is doubly logarithmic coordinates for stretched trajectories above the semicircle; (b) Comparison of the distribution $Q(r)$ with $\mathrm{Ai}^2(a_1+\ell r)$ for the radius of a semicircle $R=100$, where $a_1\approx -2.3381$ is the first zero of $\mathrm{Ai}$ and $\ell\approx 0.0811$.}
\label{fig:05}
\end{figure}

Recall that we are interested in \emph{stretched} trajectories only, meaning that we should impose the condition $t=cR$ and consider the typical width, $d(R)$ of the distribution $Q(r,R)$, where $d^2(R)$ is defined as follows:
\be
\la d^2(R)\ra = \int_{0}^{\infty} r^2\, Q\left(r, \phi=\frac{\pi}{2}, c R\right) dr -\left(\int_{0}^{\infty} r\, Q\left(r, \phi=\frac{\pi}{2}, c R\right) dr\right)^2
\label{eq:s6}
\ee
at large $R$. By the condition $t=cR$ to deal with stretched trajectories, our consideration differs from the standard diffusion process above the impenetrable disc, which was exhaustively discussed in many papers, for example, in \cite{frish}. In the figure \fig{fig:05} we have plotted (for $D=1$):
\begin{itemize}
\item[(a)] The expectation $\bar{d}(R)=\sqrt{\la d^2(R) \ra}$ as a function of $R$ in doubly logarithmic coordinates which enables us to extract the critical exponent $\gamma$ in the dependence $\bar{d}(R) \sim R^{\gamma}$ (\fig{fig:05}a),
\item[(b)] The distribution function $Q\left(r, \phi=\frac{\pi}{2}, c R\right)$ of $r$ at some fixed $c$ ($c=5$) and $R$ in comparison with the function $b\, \mathrm{Ai}^2(a_1+\ell r)$, where $\mathrm{Ai}(z)=\frac{1}{\pi} \int_{0}^{\infty} \cos(\xi^3/3+\xi z)\, d\xi$ is the Airy function (see, for example, \cite{Airy}), $a_1\approx -2.3381$ is the first zero of $\mathrm{Ai}$, $b= \left[\int_0^{\infty}\mathrm{Ai}^2(a_1+\ell r)dr\right]^{-1}$, and $\ell(c)$ is the $c$-dependent numeric constant (\fig{fig:05}b).
\end{itemize}

As one sees from \fig{fig:05}, the function $\mathrm{Ai}^2(a_1+\ell r)$ perfectly matches the probability distribution $Q\left(r, \phi=\frac{\pi}{2}, c R\right)$. The detailed analysis of this correspondence is postponed to the paper \cite{vladimirov}, which will be devoted to the discussion of the statistics of closed stretched random "flights" above the circle.

\subsection{Triangle}

The statistics of random paths above the triangle can be treated in polar coordinates centered at the tip $C$ of the triangle as shown in \fig{fig:01}b. The random walk is free in the outer sector $ACB$ with the angle $\frac{3\pi}{2}$ and zero boundary conditions at the sides $AC$ and $BC$ are applied. Seeking the solution for the corresponding diffusion equation in the form $P(r,v,t) = T(t){\cal P}(r,v)$, we have:
\be
\begin{cases}
\disp \nu^2 {\cal P}(r,v) + \left(\partial^2_{rr} + \frac{\partial_r}{r} + \frac{\partial^2_{vv}}{r^2}\right){\cal P}(r,v) = 0 \medskip \\
\disp {\cal P}(r=0,v)={\cal P}(r\rightarrow\infty, v)={\cal P}\left(r,0\right)={\cal P}\left(r,\frac{3\pi}{2}\right)=0 \medskip \\ \disp \partial_t T(t) + \nu^2 D T(t) = 0
\label{t1}
\end{cases}
\ee
Separating variables, we can write ${\cal P}(r,v)= Q(r)V(v)$ and get a set of coupled eigenvalue problems for the "angular", $v$, and "radial", $r$, coordinates.
\be
\begin{cases}
\disp \partial^2_{vv}V(v) + \lambda_n^2 V(v) = 0 \medskip \\
V\left(0\right)=V\left(\frac{3\pi}{2}\right)=0
\end{cases}; \qquad \begin{cases}
\disp \left(r^2\partial^2_{rr} +r\partial_r+ \left(\nu^2r^2-\lambda_n^2\right)\right)Q(r)=0 \medskip \\ Q(r=0)=Q(r\rightarrow\infty)=0
\end{cases}
\ee
The particular solutions to the "angular" and "radial" boundary problems read as follows:
\be
\begin{cases}
V_n\propto\sin\left(\frac{2n v}{3}\right) \medskip \\
Q_{n}\propto J_{\frac{2n}{3}}(\nu r)
\end{cases}
\label{t2-1}
\ee
The function $P(r,v,t)$ can be written now as follows:
\be
P(r,v,t)=\sum\limits_{n=1}^\infty\int\limits_{0}^\infty A_n(\nu) J_{\frac{2n}{3}}(\nu r) \sin\left(\frac{2n v}{3}\right) e^{-\nu^2 D t} d\nu
\label{t4}
\ee
where constants $A_n(\nu)$ satisfy the initial conditions:
\be
\sum\limits_{n=1}^\infty\int\limits_{0}^\infty A_n(\nu) J_{\frac{2n}{3}}(\nu r) \sin\left(\frac{2n v}{3}\right) d\nu=\delta(r-R)\delta(v-v_0)
\label{tic1}
\ee
and
\be
A_n(\nu)=\frac{4R}{3\pi}\sin\left(\frac{2nv_0}{3}\right)\nu J_{\frac{2n}{3}}(\nu R)
\label{tic2}
\ee
Rewrite the sum in (\ref{t4}) as follows:
\be
P(r,v,t)=\sum\limits_{n=1}^\infty\frac{4R}{3\pi} \sin\left(\frac{2nv_0}{3}\right)\sin\left(\frac{2nv}{3}\right)\int\limits_{0}^\infty \nu J_{\frac{2n}{3}}(\nu R) J_{\frac{2n}{3}}(\nu r)  e^{-\nu^2 D t} d\nu
\label{t7a}
\ee
Evaluating the integral in \eq{t7a}:
\be
\int\limits_{0}^\infty \nu J_{\frac{2n}{3}}(\nu R) J_{\frac{2n}{3}}(\nu r)  e^{-\nu^2 D t} d\nu=\frac{1}{2Dt}e^{-\frac{r^2+R^2}{4Dt}}I_{\frac{2n}{3}}\left(\frac{r R}{2Dt}\right)
\label{t7int}
\ee
we arrive finally at the following expression for the probability distribution:
\be
P(r,v,t)=\frac{4R}{3\pi}\frac{1}{2Dt}e^{-\frac{r^2+R^2}{4Dt}} \sum\limits_{n=1}^\infty\sin\left(\frac{2nv_0}{3}\right) \sin\left(\frac{2nv}{3}\right)I_{\frac{2n}{3}}\left(\frac{r R}{2Dt}\right)
\label{t8}
\ee

Consider a conditional probability distribution for the trajectory passing from $A$ to $B$ above the triangle through the point $D$:
\be
P(A\to D \to B)=\frac{P(A\to D)P(B\to D)}{\disp \int_{0}^\infty P(A\to D)P(B\to D) dr}
\label{tProb}
\ee
where $P(X\to D)$ is the probability to run from the point $X$ to the point $D(d,\frac{3\pi}{4})$ above the tip of the triangle. The sum in (\ref{t7int}) has the following asymptotic behavior
\be
\sum\limits_{n=1}^\infty\sin\left(\frac{2n v_0}{3}\right) \sin\left(\frac{n\pi}{2}\right)I_{\frac{2n}{3}}\left(x\right) \sim xe^{-x^{6/7}}
\label{tasymp}
\ee
Collecting \eq{t8}--\eq{tasymp}, we find the behavior of $\la d \ra$ for $t=c R$
\be
\la d \ra=\int_{0}^{\infty}r P(A\to D \to B) dr \sim \mathrm{const}
\label{tmean}
\ee
which means that the fluctuations of stretched trajectories above the tip $C$ of the triangle are bounded and do not depend on $R$. This result supports the simple scaling consideration exposed in Section II.

\section{Biased 2D random walks in a channel with forbidden voids}

\begin{figure}[ht]
\centerline{\includegraphics[width=14cm]{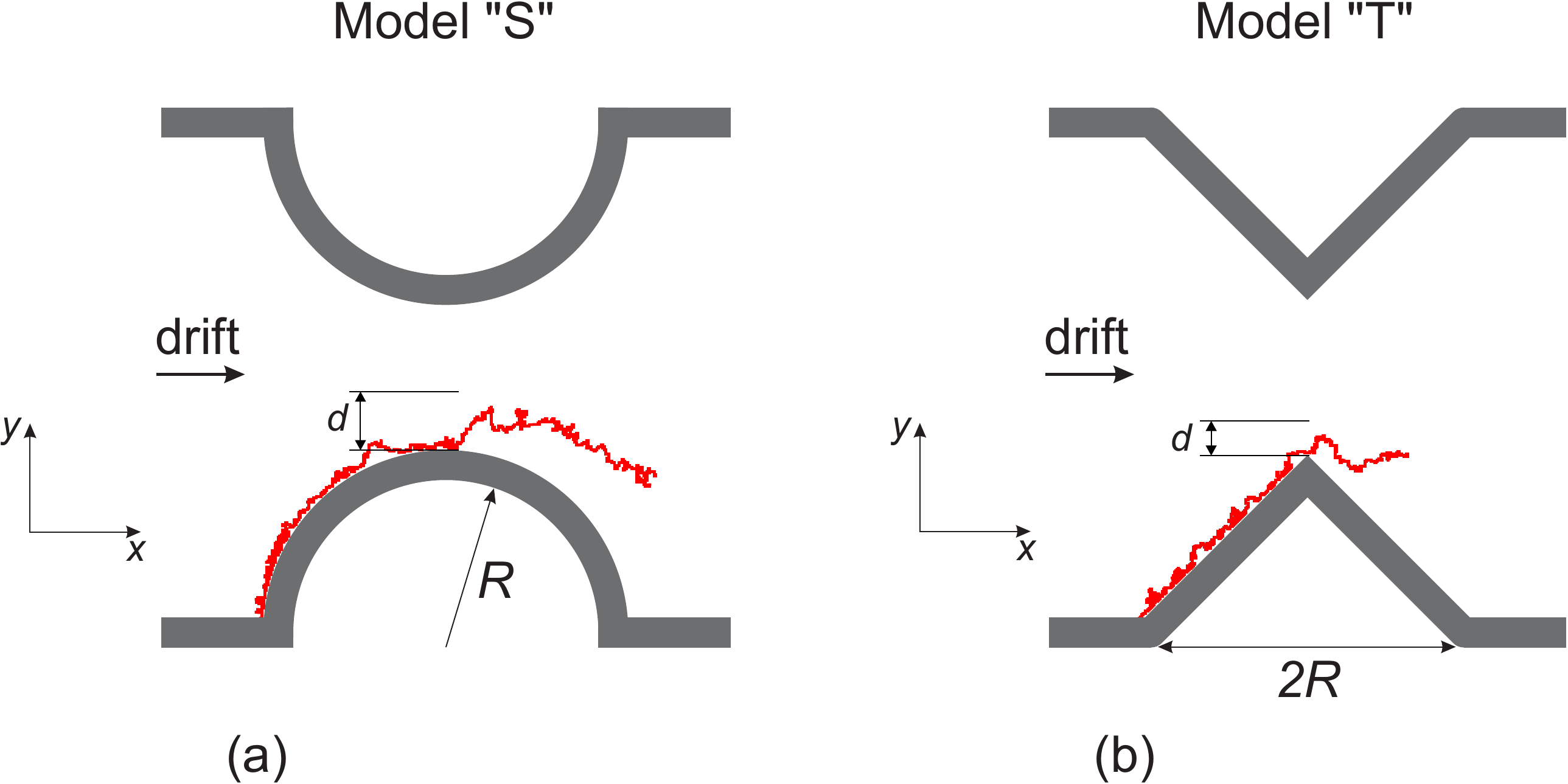}}
\caption{Biased 2D random walk in a channel with forbidden voids in a form of semicircle (a) and a triangle (b).}
\label{fig:06}
\end{figure}

As a further development of the problem of 2D random walk statistics above the semicircle and triangle, we numerically consider an ensemble of 2D random walks with a horizontal drift in a presence of forbidden voids of different shapes, as it is shown in \fig{fig:06}. The setting of this model slightly differs from the one discussed above. We regard an ensemble of long trajectories ($t\gg R$) starting at the point $A$ located to the left from the semicircle of the triangle, however we do not fix the terminal point of the path, allowing it to be everywhere. Instead of controlling the lengths of the path, $t$, we have fixed the value of the horizontal drift, $\eps$. Thus, the coordinates of the tadpole of a growing lattice path obey the following recursive transformations:
\be
(x_{t+1}, y_{t+1}) = \begin{cases} (x_t-1, y_t) & \mbox{with probability $\frac{1}{4}-\eps$} \medskip \\
(x_t+1, y_t) & \mbox{with probability $\frac{1}{4}+\frac{\eps}{3}$} \medskip \\ (x_t, y_t+1) & \mbox{with probability $\frac{1}{4}+\frac{\eps}{3}$} \medskip \\ (x_t, y_t-1) & \mbox{with probability $\frac{1}{4}+\frac{\eps}{3}$} \end{cases}
\ee
at $\eps=0$ we return to the symmetric two-dimensional random walk, while at $\eps = \frac{1}{4}$ the backward steps are completely forbidden.

\begin{figure}[ht]
\centerline{\includegraphics[width=16cm]{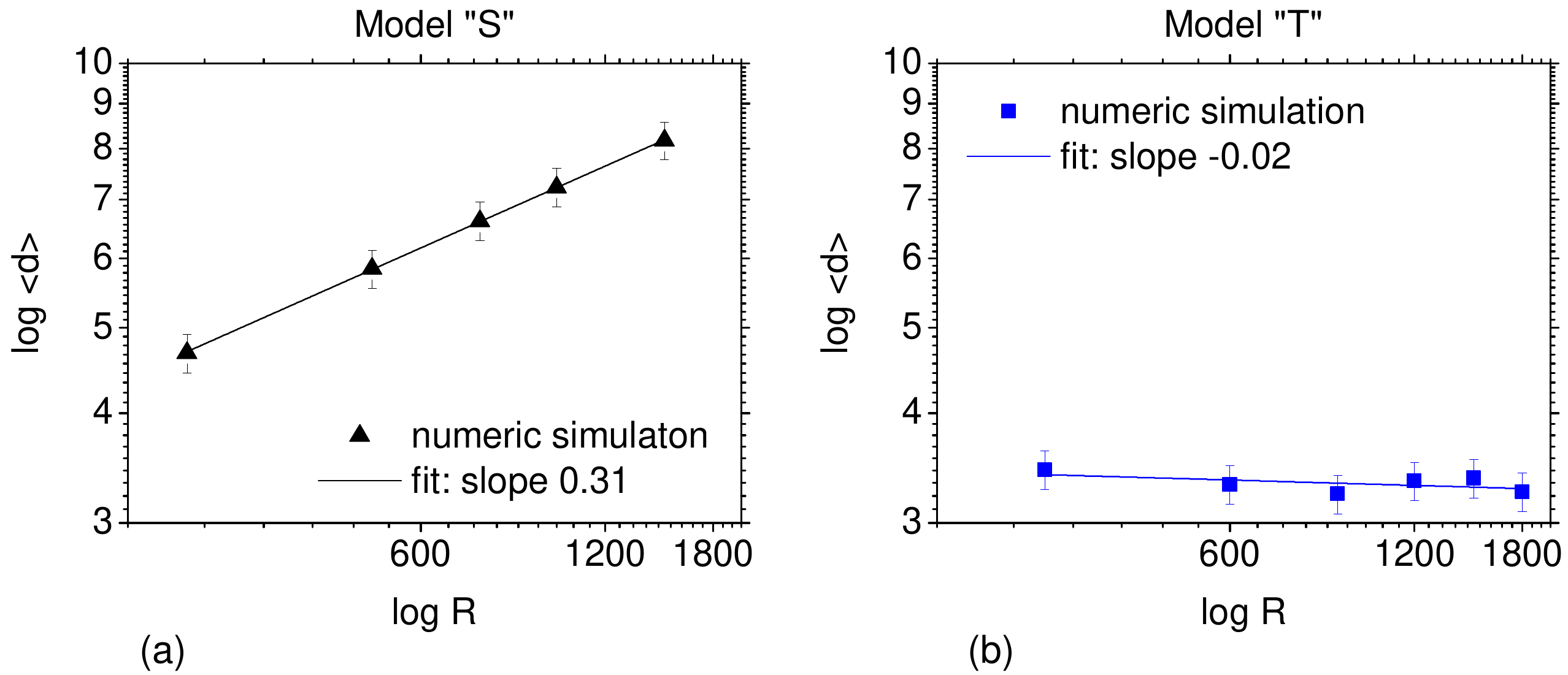}}
\caption{Mean deviations of open random paths shown in \fig{fig:06} for $\eps=\frac{3}{28}$: (a) above the top of the semicircle; (b) above the tip of the triangle.}
\label{fig:07}
\end{figure}

We have performed Monte-Carlo simulations to determine the fluctuations of 2D trajectories with the drift $\eps$ ($\eps\ge 0$) above the top of the semicircle (triangle). The corresponding results are presented in \fig{fig:07} for $\eps=\frac{3}{28}$, for which the quotient of forward to backward horizontal jump rates is equal to 2. In the case of a semicircle, the KPZ scaling for the expectation, $\la d(R)\ra \sim R^{1/3}$, holds, while for the case of the triangle the fluctuations do not depend on $R$, and the behavior $\la d(R)\ra = \mathrm{const}$ is clearly seen. We have simulated of order of $10^3$ lattice trajectories up to the length $t_{max} = 2\times 10^3$ in the presence of voids characterized by $R=\{250, 500, 750, 1000, 1250, 1500\}$ (measured in the units of lattice spacing). Thus, the statistics of biased 2D random walks in presence of forbidden voids of semicircular and triangular shapes matches the fluctuations of stretched 2D random walks above the same shapes discussed at length of the Section II.

Found behavior of biased random walks in vicinity of excluded voids of various shapes, allows us to make a conjecture about possible thermodynamic properties of laminar flows in tubes with periodic contractions. The combination of the drift and geometry pushes the laminar flow lines which spread near the boundary, into a large deviation regime with the extreme value statistics, typical for 1D systems with spatial correlations. Since the width of the fluctuational (skin) layer near the boundary is shape-dependent, one may expect different heat emission of laminar flows in presence of excluded voids of different geometries.

\section{Discussion}

In this work we considered simple two-dimensional systems in which imposed external constraints push the underlying stochastic processes into the "improbable" (i.e. large deviation) regime possessing the anomalous statistics. Specifically, we dealt with the fluctuations of a two-dimensional random walk above the semicircle and the triangle in a special case of "stretched" trajectories. We proposed the simple scaling arguments supported by the analytic consideration. As a brief outline of the results, it is worth highlighting three important points:
\begin{itemize}
\item Imposing constraints on a conformational space, which cut off a tiny region of available ensemble of trajectories, we can push the sub-ensemble of random walks into the atypical large deviation regime possessing anomalous fluctuations, which could have some similarities with the statistics of correlated random variables;
\item Stretching 2D random paths above the semicircle, we may effectively reduce the space dimension: in specific geometries we force the system to display the 1D KPZ fluctuations;
\item Strong dependence of the fluctuation exponent $\gamma$ on the geometry of the excluded area, manifests the non-universality in the underlying reduction of the dimension. We outline three archetypical geometries: stretching above the plane (Gaussian, with $\gamma=1/2$), above the semicircle (KPZ-type, with $\gamma=1/3$) and above the triangle or the cusp (finite, with $\gamma=0$). For an algebraic curve of order $\eta$ the fluctuation exponent is $\gamma = \frac{\eta-1}{2\eta-1}$.
\end{itemize}

Our results demonstrate that geometry has a crucial impact on the width of the boundary layer in which the laminar flow lines diffuse. We could speculate that such an effect is important for some technical applications in rheology of viscous liquids, for instance, for cooling of laminar flows in channels with periodically displaced excluded voids of various shapes (like shown in \fig{fig:05}). Such a conjecture is based on the following obvious fact. The heat transfer through walls depends not only on the total contact surface of the flow with the wall, but also on a width of a mixing skin layer: the bigger a mixing layer near the boundary, the better cooling. However as we have seen throughout the paper, the width of the mixing layer is shape-dependent, and hence, it might control the "optimal" channel geometry for cooling of laminar viscous liquids flows.

\begin{figure}[ht]
\centerline{\includegraphics[width=14cm]{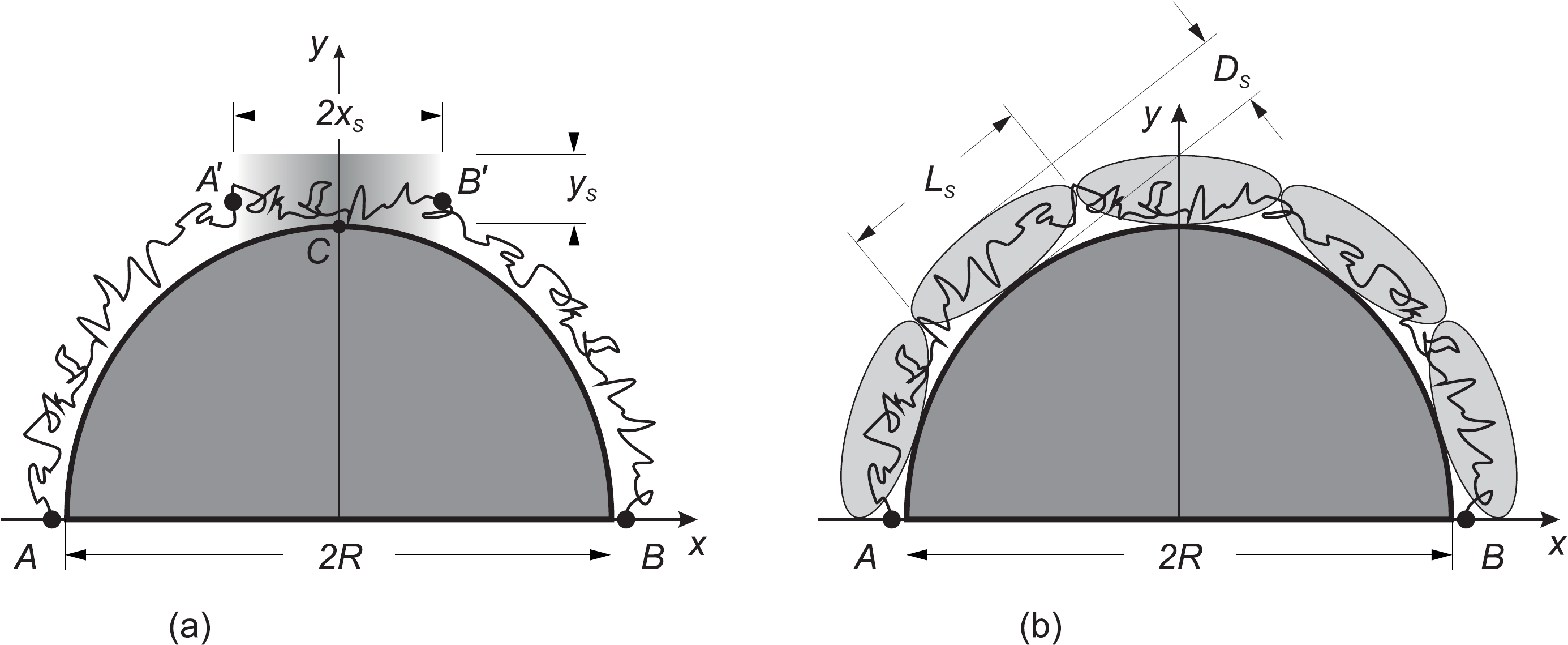}}
\caption{(a) Two-dimensional random walk evading the semicircle. The part $A'B'$ lies above the essentially flat region of the semicircle; (b) splitting in blobs of a trajectory evading curved surface (semicircle).}
\label{fig:08}
\end{figure}

The 1D KPZ-type behavior in a 2D restricted random walk goes far beyond the pure academic interest. Two important relevant applications should be mentioned. First, by this model we provide an explicit example of the two-dimensional statistical system which, being pushed to the large-deviation ("atypical") region, mimics the behavior of some one-dimensional correlated stochastic process. Second, our study deals with the manifestations of a 1D KPZ-type scaling in the localization phenomena of 2D constrained disordered systems. Namely, let us estimate the free energy, $F(N)$ of an ensemble of $N$-step paths stretched above the semicircle as shown in \fig{fig:08}a.

One can split the entire stretched path of length $N$ running from $A$ to $B$ above the semicircle into the sequence of independent "blobs" with the longitudinal size $L_S=x_S \sim R^{2/3}$ and the transversal size $D_S=y_s\sim R^{1/3}$ -- see \fig{fig:08}b. Thus, taking into account the additive character of the free energy, we can estimate $F(N)$ of ensemble of $N=cR$--step paths as
\be
F(R) \sim \frac{N}{L_S} \sim \frac{R}{R^{2/3}} \sim R^{1/3}
\label{eq:04}
\ee
Therefore, the Gibs measure, which provides expression of the "survival probability" in the curved channel of length $N \sim R$ and diameter $\sim R^{1/3}$, can be estimated as follows
\be
P(R) = e^{-F(R)} \sim e^{-\alpha R^{1/3}}
\label{eq:05}
\ee
where $\alpha$ is some model-dependent numerical constant. Passing to the grand canonical formulation of the problem, i.e. attributing the energy $E$ to each step of the path (remembering that $N = cR$), one can rewrite the expression for $P(R)$ in \eq{eq:05} as follows
\be
P(E) = \int_0^{\infty} P(R)\, e^{-ER}\, dR \sim \varphi(E)\, e^{-b/\sqrt{E}}
\label{eq:05a}
\ee
where $b=\frac{2\alpha^{3/2}}{3^{3/2}}$ and $\varphi(E)$ is a power-law function of $E$.

To provide some speculations behind the behavior \eq{eq:05a}, recall that the density of states, $r(E)$, of the 1D Anderson model (the tight-binding model with the randomness on the main diagonal) at $E\to 0$, has the asymptotics \eq{eq:05a}, known as the "Lifshitz singularity", $r(E) \sim e^{-a/\sqrt{E}}$, where $E$ is the energy of the system and $a$ is some positive constant (see \cite{lif1,lif2} for more details).

The asymptotics \eq{eq:05}, has appeared in the literature under various names, like "stretched exponent", "Griffiths singularity", "Balagurov-Waks trapping exponent", however, as mentioned in \cite{neuwen}, in all cases this is nothing else as the inverse Laplace-transformed Lifshitz tail of the one-dimensional disordered systems possessing Anderson localization \eq{eq:05a}. We claim that the KPZ-type behavior with the critical exponent $\gamma=\frac{1}{3}$ can also be regarded as an incarnation of a specific "optimal fluctuation in a large deviations regime" for the one-dimensional Anderson localization. Finding in some 2D systems a behavior typical for 1D localization, seems to be a challenging problem of connecting localization in constrained 2D and 1D systems. In details this issue will be discussed in a forthcoming publication.

\begin{acknowledgments}

We are grateful to V. Avetisov, A. Gorsky, A. Grosberg, B. Meerson, S. Pirogov and M. Tamm for number of fruitful discussions and useful critical remarks. The work of S.N. is partially supported by the RFBR grant No. 16-02-00252; K.P. acknowledges the support of the Foundation for the Support of Theoretical Physics and Mathematics "BASIS" (grant 17-12-278); The work of A. Vladimirov was supported by RFBR grant 16-29-09497. The work of A. Valov was performed within frameworks of the state task for ICP RAS 0082-2014-0001 (registration \#AAAA-A17-117040610310-6). The work of S.S. was supported by the RFBR--CNRS grant No. 17-51-150006.

\end{acknowledgments}

\end{document}